\documentclass[journal,article,submit,pdftex,moreauthors]{Definitions/mdpi} 
\preto{\abstractkeywords}{\nolinenumbers} 
\usepackage{graphicx}
\usepackage{subcaption}
\firstpage{1} 
\pubvolume{1}
\issuenum{1}
\articlenumber{0}
\pubyear{2024}
\copyrightyear{2024}
\datereceived{ } 
\daterevised{ } % Comment out if no revised date
\dateaccepted{ } 
\datepublished{ } 
\hreflink{https://doi.org/} % If needed use \linebreak
\Title{Exploring the Premelting Transition through Molecular Simulations Powered by Neural Network Potentials}

\TitleCitation{Title}

\Author{Limin Zeng $^{1}$, and Ang Gao $^{1}$*}

\AuthorNames{Limin Zeng and Ang Gao}

\AuthorCitation{Zeng, L.; Gao, A.}
\address[1]{%
$^{1}$ \quad School of Science, Beijing University of Posts and Telecommunications}

\corres{Correspondence: anggao@bupt.edu.cn; }

\abstract{The premelting layer at the ice-air interface plays critical roles in atmospheric chemical and physical processes. Traditional approaches to study this layer—experimental methods, classical simulations, and even first-principles simulations—face significant limitations in accuracy and scalability. Addressing these challenges, we employ molecular dynamics simulations based on neural network potentials, offering accuracy comparable to first-principles calculations but with much higher computational efficiency. This advancement enables us to simulate the ice-air interface at a much larger scale. Our simulations reveal that the number of melted ice layers at the interface increases logarithmically with temperature, and the transition temperatures for complete melting surpass the bulk ice melting temperature, contrary to Lifshitz theory. Additionally, our findings unveil a significant dynamic heterogeneity within the premelting layer, akin to that observed in supercooled liquids, thereby substantiating the hypothesis that the premelting layer encompasses both ice-like and liquid-like phases.}


\begin{document}

%%%%%%%%%%%%%%%%%%%%%%%%%%%%%%%%%%%%%%%%%%
% The order of the section titles is different for some journals. Please refer to the "Instructions for Authors” on the journal homepage.

\section{Introduction}

The surface of ice in contact with air is covered by a thin film of liquid water, known as the premelting layer. Initially proposed by Faraday in 1842 as an explanation for the reduced friction of ice surfaces\cite{Faraday1859}, the existence of the premelting layer was not empirically confirmed until 1987\cite{KOUCHI1987}. Subsequent researches have demonstrated that premelting layers are not unique to ice but also occur on various other crystalline surfaces\cite{Pluis1989}.

The premelting layer plays a crucial role in mediating physical and chemical processes within the atmosphere. Notably, it facilitates the transfer of electrical charges between colliding ice particles in clouds, a mechanism instrumental in lightning formation\cite{Dash2001}. Additionally, due to its enhanced solubility for ions, the premelting layer serves as a key locus for a variety of atmospheric chemical reactions. For instance, within this layer, hypochlorous acid undergoes conversion to chlorine gas, which subsequently contributes to ozone depletion upon its release into the atmosphere\cite{Bianco2006}. Similarly, the transformation of atmospheric sulfur dioxide into sulfuric acid within the premelting layer is a significant factor in the acidification of snow\cite{Abbatt2003, Conklin1993}.

Recent progress in the field, underpinned by computer simulations and experimental findings, revealed the  heterogeneous nature of the premelting layer, indicating it may consist of multiple thermodynamic phases. Computer simulation based studies demonstrate that the layer is not a simple uniform film of liquid water but contains regions that exhibit both liquid-like and solid-like properties, with the ability to transition dynamically between these states\cite{Hudait2017, Pickering2018}.  Recent experiments show that  many droplets emerge on the surface of the premelting layer during the growth of ice crystals\cite{Sazaki2012, Sazaki2013, Asakawa2015}. This has led to the hypothesis that these droplets may represent a distinct thermodynamic phase\cite{Sazaki2012, Sazaki2013, Asakawa2015, Murata2016}, suggesting a complex structural composition of the premelting layer.
 
Despite the progress made in understanding the premelting layer, significant uncertainties remain regarding the microscopic mechanisms that dictate its structural and dynamic characteristics. The variation in thickness measurements obtained through different experimental techniques underscores the challenges in accurately characterizing this complex interfacial layer\cite{KOUCHI1987, Dosch1995, Bluhm2002}. Theoretical explanations for the premelting layer have largely been based on Lifshitz theory\cite{Dash2006, Dzyaloshinskii1961}, which emphasizes the role of van der Waals forces but tends to neglect the critical role played by hydrogen bonds between water molecules. While Lifshitz theory has offered insights into the appearance of droplets on the premelting layer during ice crystal growth\cite{Sibley2021}, the theory's quantitative predictions have yet to be confirmed by experimental evidence\cite{Limmer2014}.

In this study, we utilize molecular dynamics simulations enhanced by neural network potentials to probe the premelting layer, aiming to surpass the limitations inherent in experimental methodologies and classical simulation algorithms. Simulations are able to provide insights into the microscopic mechanisms governing the premelting layer, which are often obscured in experimental studies. Traditionally, the field has relied on classical simulation algorithms\cite{Hudait2017, Pickering2018,Limmer2014,Niblett2021}, which are constrained by the use of artificially constructed force fields and the challenges in capturing complex effects such as breaking of chemical bonds, and electron polarization effects.

Transitioning to first-principles molecular dynamics simulations addresses these constraints, delivering results that markedly differ from classical approaches \cite{Paesani2008, Watkins2011}. However, the computational intensity of first-principles simulations limits their application to small-sized systems, thus constraining our ability to draw comprehensive conclusions about the premelting layer's heterogeneity or thickness near the melting point.

To overcome these limitations, we utilize machine learning-based first-principles simulations, significantly enhancing computational efficiency. Neural network based potentials, trained on data derived from density functional theory (DFT), is able to not only achieve a level of accuracy on par with the DFT calculations itself\cite{Behler2011, Behler:2021aa, zhang2021phase, Gao:2022ws} but also operate at a significantly higher computational speed\cite{Mailoa:2019va}. This approach enables us to extend the time and spatial scales of our simulations beyond the reach of conventional density functional theory-based methods, allowing for a more detailed and accurate exploration of the premelting layer while maintaining the rigor of first-principles accuracy.

Utilizing the neural network potentials detailed earlier, this work conducts large-scale molecular dynamics simulations to explore the premelting layer at the air-ice interface across varying temperatures. The results indicate that the number of melted ice layers increases logarithmically with temperature, albeit with slight variations observed across the basal, prismatic, and secondary prismatic planes. Contrary to Lifshitz theory's predictions\cite{Dash2006, Dzyaloshinskii1961}, the transition temperatures for complete melting observed in our simulations exceed the bulk ice melting temperature. Furthermore, we observed dynamic heterogeneity within the premelting layer, reminiscent of the behavior seen in supercooled liquids\cite{10.1063/1.1992481,doi:10.1126/science.abb9385,doi:10.1126/science.abb9796,Palmer:2014wr,Sastry:2003wb}, supporting the hypothesis that the premelting layer consists of both liquid-like and ice-like regions.

%%%%%%%%%%%%%%%%%%%%%%%%%%%%%%%%%%%%%%%%%%
\section{Materials and Methods}

\subsection{Simulation Settings}
In this work, we have employed the neural network potential for water molecules developed by Wohlfahrt et al \cite{DellagoJCP}. This potential is derived from a comprehensive dataset consisting of 8007 configurations,
encompassing a broad spectrum of water states including bulk liquid, bulk ice, water-air, and ice-water interfaces. The energy and forces of these configurations are obtained by DFT calculations with the RevPBE functional\cite{PhysRevB.59.7413} augmented by Grimme's D3 correction\cite{10.1063/1.3382344}. These training configurations are processed by the n2p2 software\cite{Singraber:2019vs,Singraber:2019wi} to generate the neural network potential. Previous studies\cite{Morawietz2016, DellagoJCP} have shown that this neural network potential can accurately predict properties such as the melting point of ice and the water-vapor coexistence curve, making it an excellent choice for studying the premelting layer. 

\begin{figure}[h]
\centering
\hspace{10pt}
\includegraphics[scale=0.048]{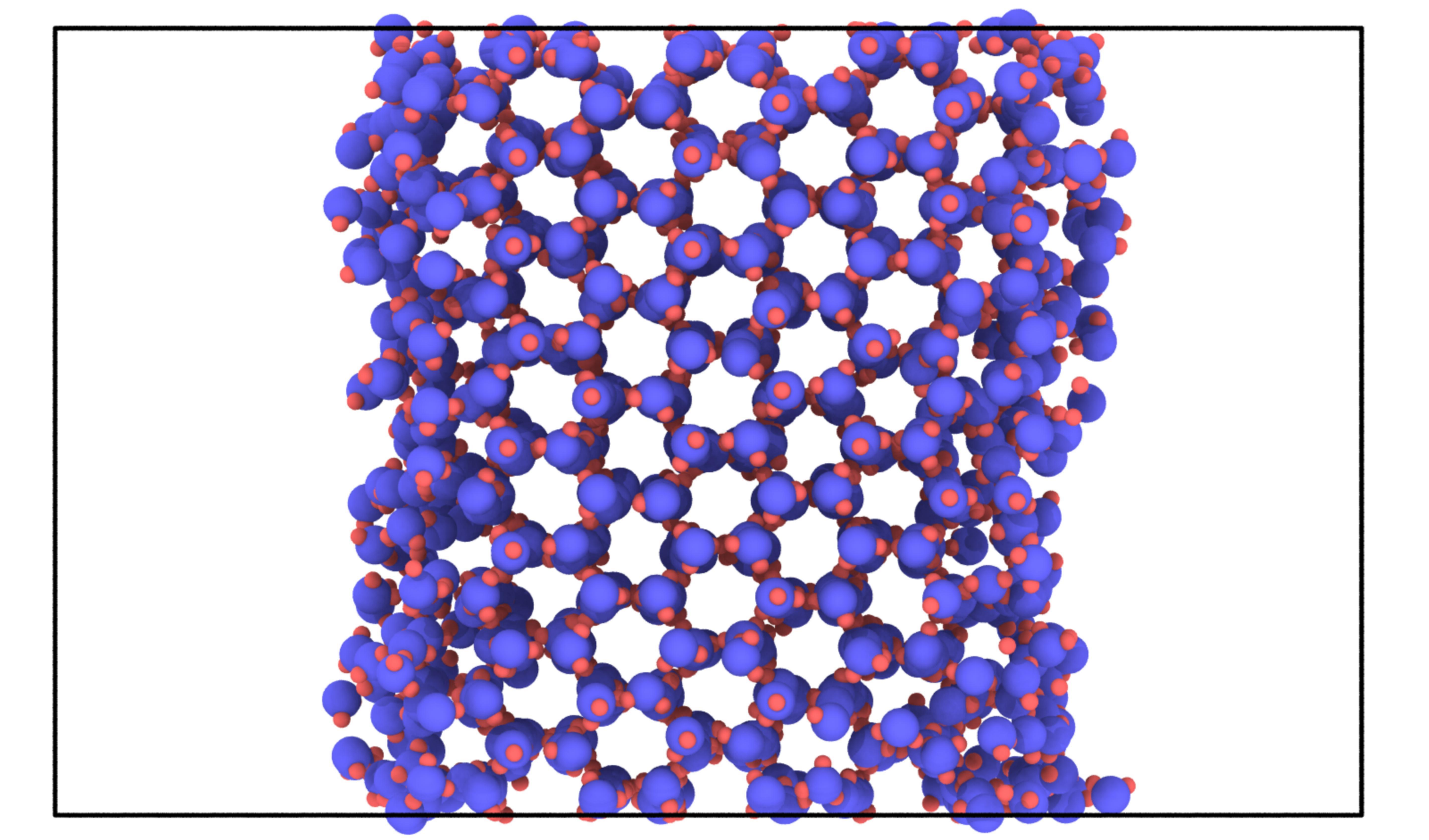}
\caption{A snapshot of the simulation box with premelting layer on the primary prismatic facets observed through the basal facets.  Without extension, in the original box which is filled with ice, the thickness of the stacking of the basal facets, primary prismatic facets and secondary prismatic facets is 2.94143 nm, 3.61461 nm and 3.12914 nm, respectively. The length of the edges along the extension direction of the simulation box is set to 6 nm while the length of the other two edges is the same as that of the original box. After extension, in the profile parallel to the extension direction, you can see air, water and ice at the same time.}
\centering
\label{fig:slab}
\end{figure}

The system we simulated consists of a slab of ice in contact with vapor on both sides, as depicted in Figure~\ref{fig:slab}. Periodic boundary conditions are employed to minimize boundary effects. We focused on ice in the Ih phase, which is the most common ice phase found in nature.  The structure of the iceIh was generated using the Genice2 software\cite{https://doi.org/10.1002/jcc.25077},  arranging the ice into a 4x4x4 supercell structure, with each unit cell comprising 16 water molecules. Consequently, the simulated system contains 1,024 water molecules, a scale significantly exceeding that of prior first-principles studies on the premelting layer. This enhanced scale allows for a more detailed exploration of the premelting layer's properties and behaviors under various conditions.

\begin{figure}[h]
\centering
\hspace{10pt}
\includegraphics[scale=0.08]{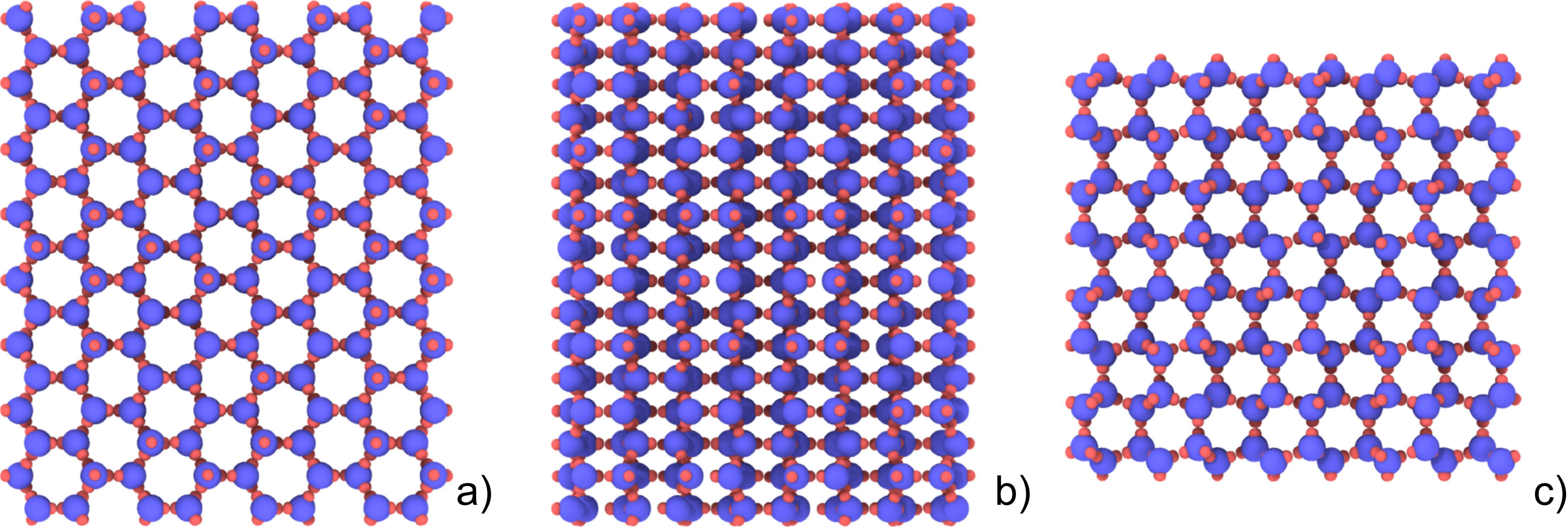}
\captionsetup{justification=centering}
\caption{From left to right: the basal, primary prismatic, and secondary prismatic facets.}
\label{fig:facets}
\centering
\end{figure}

The most common crystal planes of iceIh  in contact with air are the basal,  primary prismatic and secondary prismatic facets of ice(Figure~\ref{fig:facets}). These different facets are known to affect the thickness of the premelting layer\cite{Dosch1995}. Therefore, in this work, we separately investigated  these specific interfaces of ice in contact with air  to understand how the unique characteristics of each facet affect the premelting layer.

The molecular dynamics simulations are conducted using the 21Nov2023 version of the LAMMPS software, which has builtin n2p2 support. The simulations are carried out in the NVT ensemble. The time step for the simulation was set to 1fs.  

\subsection{Number of Melted Layers}
In this work, we use the number of melted ice layers  as the order parameter of the premelting transition. To determine whether a layer is melted, we rely on the density profile of oxygen atoms at the air-ice boundary, formulated as:
\begin{equation}
\rho(z) = \frac{1}{N} \left\langle\sum_{i=1}^N \delta(z_i - z)\right\rangle \,,
\end{equation}
where \(N\) denotes the total number of water molecules in the system, \(z_i\) is the z-coordinate of the Oxygen atom of the \(i\)-th molecule , and \(\left\langle\cdots\right\rangle\) indicates the ensemble average.

We call an ice layer melted if the peak-valley structure of that layer is no longer distinct on the density profile. Or more quantitatively, if the height of the peak corresponding to that layer is less than twice the height of its neighbouring valley, we call that layer melted. 

\subsection{Propensity}
In this study, we explore the dynamical properties of the premelting layer by examining the molecular propensity within it. Propensity, a well-established metric, gauges the mobility of molecules and requires simulations within the isoconfigurational ensemble\cite{10.1063/1.2719192, PhysRevE.76.041509}. This ensemble comprises trajectories that, while originating from the same structural configuration, diverge due to initially velocities being sampled from a Maxwell distribution. Through this ensemble, we can calculate the dynamical properties of molecules, including diffusion distances, by averaging over these varied paths. The focus of our analysis is on the propensity\cite{10.1063/1.2719192, PhysRevE.76.041509}, defined as the average squared displacement of each molecule:
\begin{equation}
\text{propensity}(i) = \langle||r_i(t) - r_i(0)||_2\rangle_{\text{iso}},
\end{equation}
where \(i\) indicates the \(i\)-th molecule, identified by its oxygen atom, \(\langle\cdots\rangle_{\text{iso}}\) averages over the isoconfigurational ensemble, \(r_i(t)\) marks the position of the molecule's oxygen atom at time \(t\), and \(||\cdots||_2\) denotes the L2 norm.

For our study, we selected configurations featuring a thin premelting layer on the basal, primary prismatic, and secondary prismatic facets. From each configuration, we simulated 10 trajectories, with the molecular propensity derived by averaging across these simulations."

%%%%%%%%%%%%%%%%%%%%%%%%%%%%%%%%%%%%%%%%%%
\section{Results}
\subsection{Premelting Transition}

To investigate the premelting transition as the temperature tends to the transition temperature, we perform a series of simulations at temperatures ranging from 175K to 310K. All the simulations are initiated  with the identical configuration where ice in its pristine  crystalline form are in contact with air.   The system undergoes at least 40 ps  simulation for equilibration, followed by an additional 50 ps of simulation to collect data essential for determining the thermodynamic properties.

\begin{figure}[h]
\hspace{27pt}
\includegraphics[scale=0.08]{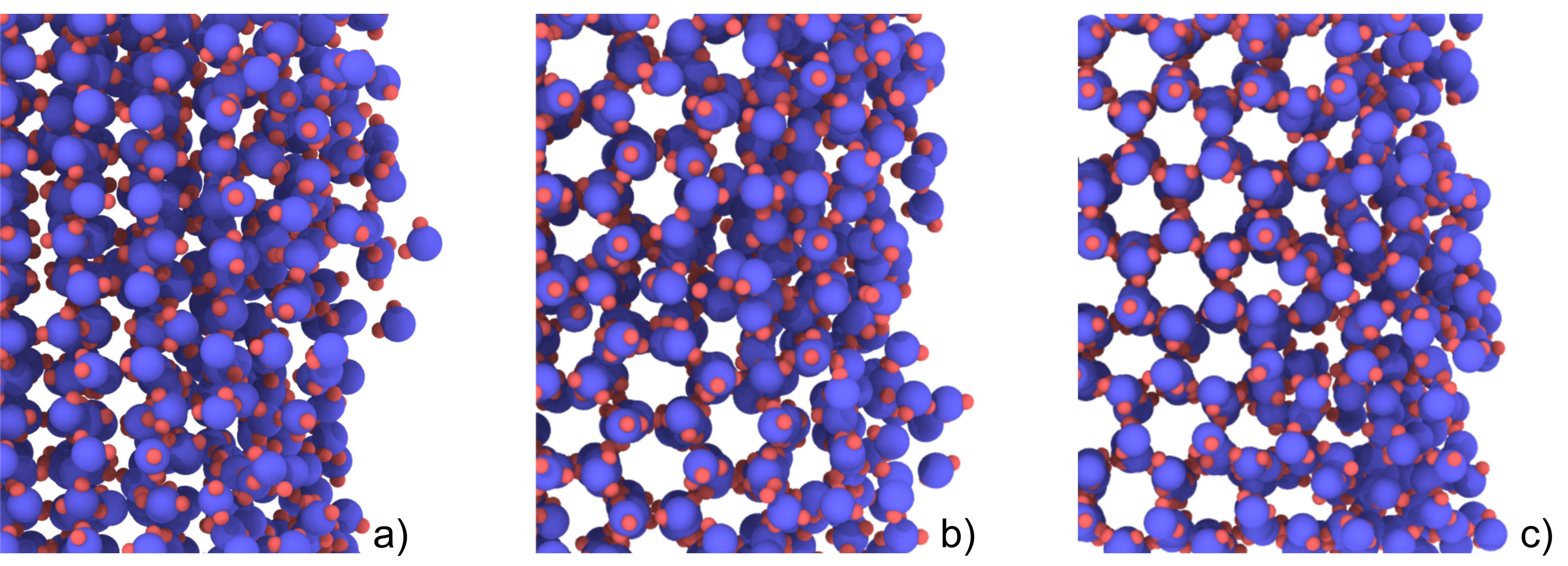}

\caption{Premelting layer on different facets at 295K. a) shows premelting layer on the basal facets observed through the primary prismatic facets. b) shows  premelting layer on the primary prismatic facets observed through the basal facets. c) shows premelting layer on the secondary prismatic facets observed through the basal facets.}
\label{fig:facets_melted}
\end{figure}

Snapshots of the configurations obtained at 295K is shown in Figure~\ref{fig:facets_melted}, where a thin premelting layer is formed on the primary prismatic, secondary prismatic, and basal facets, respectively.

\begin{figure}[h]
\hspace{65pt}
\includegraphics[scale=0.14]{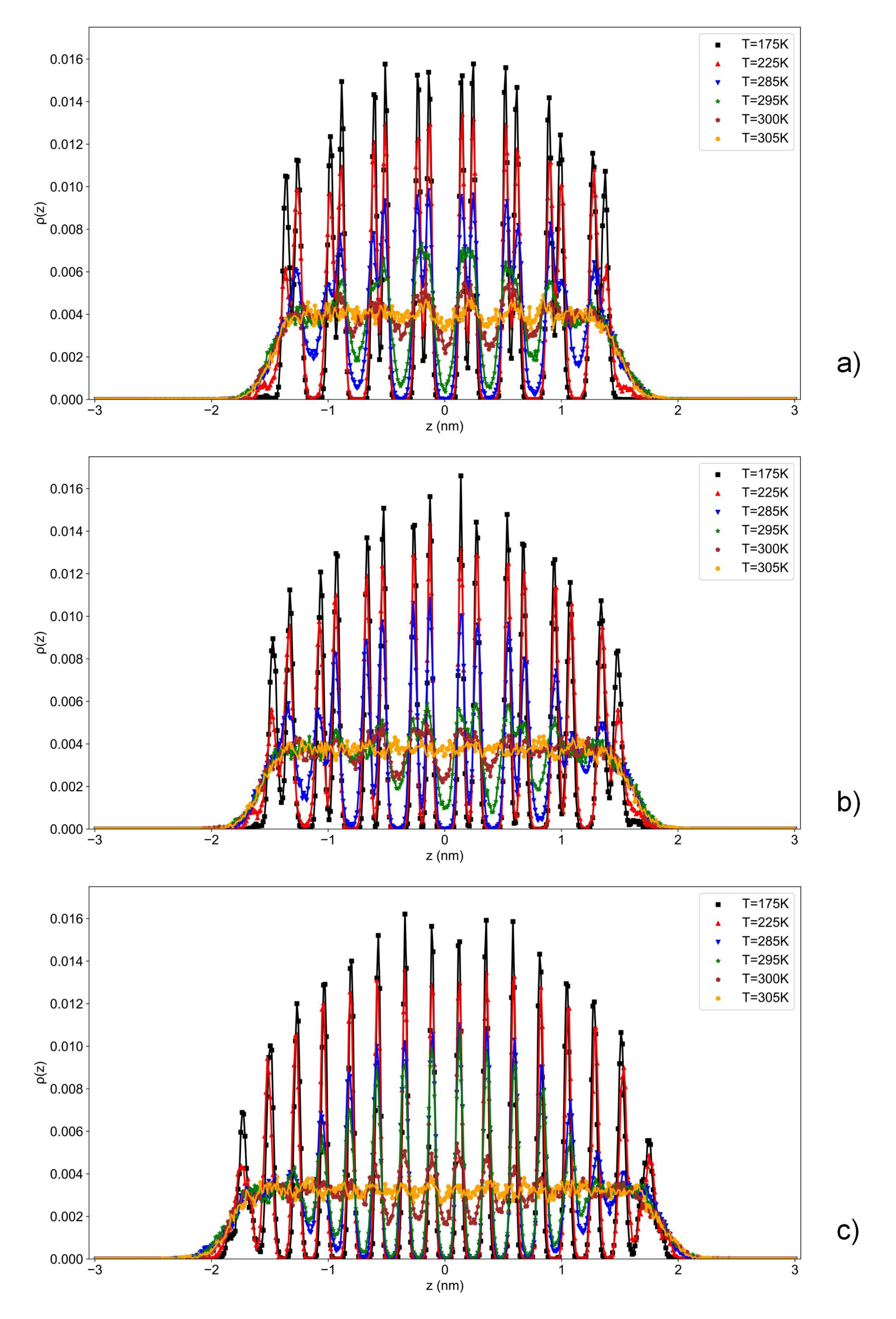}
\caption{From top to bottom: density profiles of basal, primary prismatic and secondary prismatic facets. Density profiles of different facets at 175K (black square connected by black line ), 225K (red upright triangle connected by red line ), 285K (blue downright triangle connected with blue line ), 295K (green pentagram connected by green line ), 300K (brown pentagon connected by brown line ), 305K (orange hexagon connected by orange line ) are shown.}
\label{fig:density_profile}
\end{figure}

Figures~\ref{fig:density_profile} display the density profiles at various temperatures for the primary prismatic, secondary prismatic, and basal facets, respectively. The illustrations reveal that at lower temperatures, the premelting layer is  is essentially absent. However, as temperature approaches the transition temperature, there is a significant increase in the number of melted ice layers, eventually leading to the complete melting of the layers.

\begin{figure}[h]
\hspace{105pt}
\includegraphics[scale=0.28]{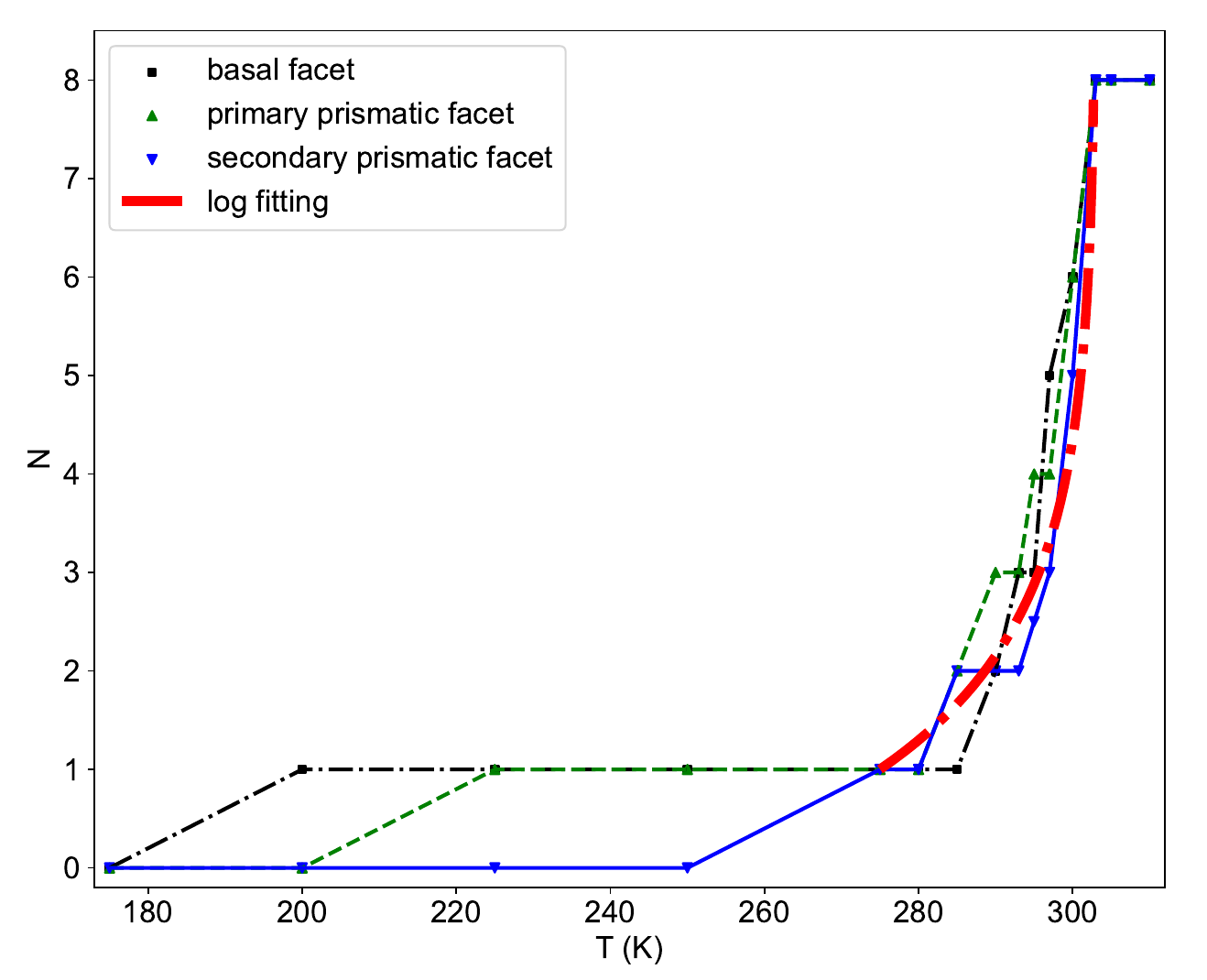}
\caption{Number of melted layers N on different facets as a function of temperature T. Black squares connected by black dash-dotted line stands for N on basal facets. Green upright triangles connected by green dashed line stands for N on primary prismatic facets. Blue downright triangles connected by blue line stands for N on secondary prismatic facets. Log fitting curve (red dash-dotted line) is a fitting of the N-T curve near the critical temperature according to Lifshitz theory.}
\label{fig:num_melted}
\end{figure}

Figure~\ref{fig:num_melted} offers a concise depiction of how the quantity of melted ice layers varies with temperature across the different facets. It is observed that the number of melted layers differs among the primary prismatic, secondary prismatic, and basal facets. This observation contradicts the Lifshitz theory, which postulates an uniform premelting layer thickness across all crystalline surfaces. The deviation may be explained by considering the Lifshitz theory's lack of consideration for the hydrogen bonding interactions among water molecules.  Moreover, the transition temperatures where all the layers are melted is about 300K, which is different from the melting temperature of bulk ice as determined by this neural network potential (273K). Such a discrepancy challenges  Lifshitz theory's assertion that the transition temperature for premelting should coincide with the melting temperature of bulk ice. The observed higher transition temperature may be due to that the unbalanced forces experienced by the particles at the interface has a strong effect on the melting behavior.

\subsection{Dynamic Heterogeneity}
Recently, Hudait et al\cite{Hudait2017} and Pickering et al\cite{Pickering2018} investigated the heterogeneity of the premelting layer through molecular dynamics simulations using the mW water model. Their research show that the premelting layer on ice is not a homogenous liquid layer; instead, it consists of regions with ice-like molecules and regions with liquid-like molecules. Moreover, they discovered that these ice-like and water-like regions are not static; they can dynamically transition between one another. However, it's important to highlight that their categorization of molecules as liquid-like or ice-like is solely based on structural characteristics. A molecule is identified as ice like if its hydrogen bondings  with neighboring molecules are similar to those in ice, and a molecule will be is identified as liquid-like if the hydrogen bondings are more like those in liquid water. The dynamic properties of the liquid is not considered in this characterization.

In this work, we aim to verify this hypothesis of the exsistence of ice-like and liquid-like regions by examining the dynamic properties of molecules in the premelted layer. The ice-like molecules are surrounded by  more intact hydrogen bonding network, which is harder to  escape and therefore the ice-like molecules should be less mobile than the liquid-like molecule. In this work, we use propensity as the metric to characterize the mobility of the molecules, which is a metric commonly used to measure the  dynamics of supercooled liquids.

\begin{figure}[h]
\hspace{12pt}
\includegraphics[scale=0.09]{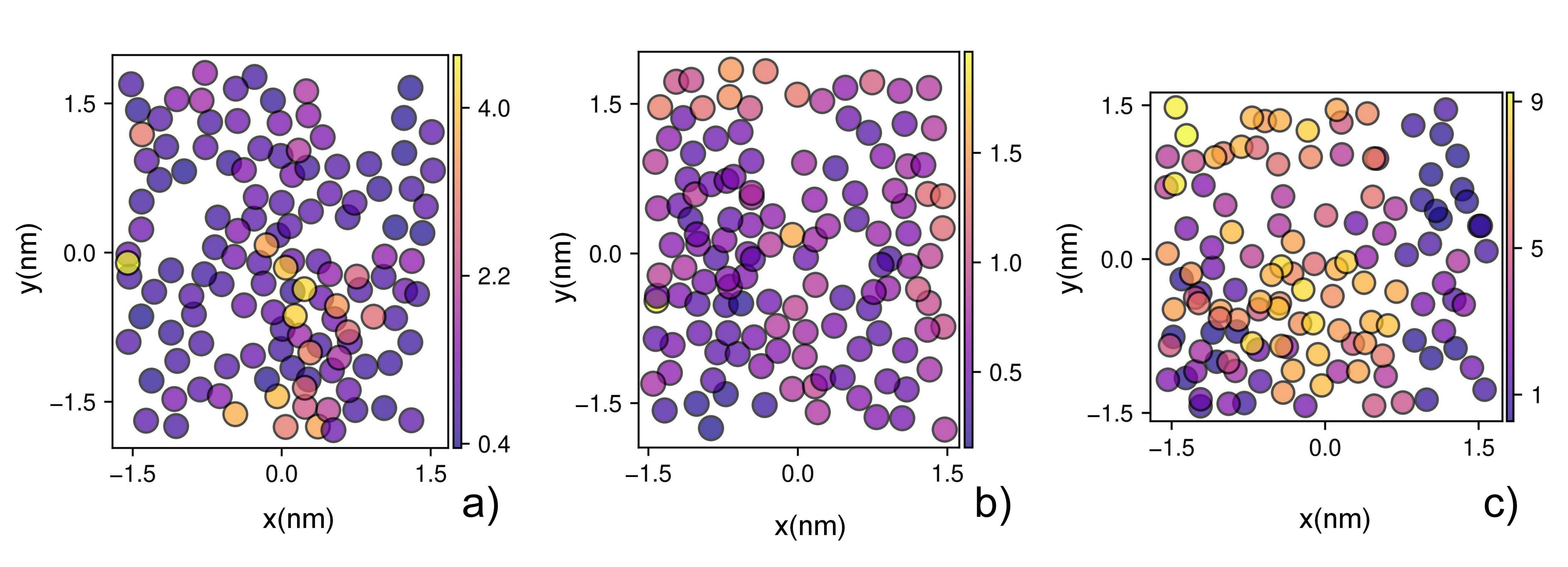}
\caption{Propensity of premelting layer at different facets.From left to right: the basal, primary prismatic, and secondary prismatic facets.}
\label{fig:propensity}
\end{figure}

The propensity of particles across the premelting layer on the basal, primary prismatic, and secondary prismatic facets is illustrated in Figure~\ref{fig:propensity}. This propensity map reveals clear regions of both high and low mobility, aligning with the hypothesis that the premelting layer comprises both liquid-like and ice-like regions. This dynamic heterogeneity highlighted by the segregation is  a determining feature of supercooled liquids. Such observations lend support to the idea that, just like the supercooled bulk water,  a liquid-liquid coexistence also occurs on the premelting layer, where the ice-like molecules corresponds to the low density liquid (LDL) phase and the liquid-like molecules corresponds to the high density liquid (HDL) phase\cite{doi:10.1073/pnas.2008426117,doi:10.1126/science.1103073,PhysRevE.62.6968}. 

%%%%%%%%%%%%%%%%%%%%%%%%%%%%%%%%%%%%%%%%%%
\section{Discussion}

Our study of the premelting transition at the ice-air interface, as unveiled through molecular dynamics simulations utilizing neural network potentials, has provided groundbreaking insights into the intricate behaviors of the premelting layer. Our discoveries challenge longstanding theories, such as Lifshitz theory, and shed light on the dynamic heterogeneity within the premelting layer, thereby echoling and extending  the findings of previous studies.

The observation that the complete melting occurs at temperatures exceeding the bulk ice melting temperature (determined to be 273K by the neural network potential we used ) raises important questions about the roles that different forces  play at the ice-air interface. This discrepancy from Lifshitz theory, which anticipates a transition temperature coinciding with bulk ice melting, suggests that factors such as hydrogen bonding and the unbalanced electrostatic forces experienced by interfacial molecules may significantly influence the melting behavior. Such insights question the adequacy of Lifshitz theory, which primarily considers van der Waals interactions, and underscore the need for theoretical models to more thoroughly integrate diverse interactions.

Furthermore, the observed dynamic heterogeneity in the premelting layer, marked by areas of varying mobility, provides compelling evidence supporting the hypothesis of the existence of ice-like regions and liquid-like regions in the premelting layer. This observation, akin to the dynamics of supercooled liquids, suggests that the premelting layer may also undergo a liquid-liquid phase transition  in a manner similar to supercooled bulk water undergoing a liquid-liquid phase transition, with ice-like regions corresponding to a low-density liquid phase and liquid-like regions to a high-density liquid phase. 

Looking forward, several directions for future research may emerge from our study. Firstly, extending the simulations to include the effects of impurities and atmospheric gases could provide a more realistic picture of the premelting dynamics in natural environments. Additionally, exploring the implications of the dynamic heterogeneity and liquid-liquid coexistence of the premelting layer for the macroscopic properties of ice, such as its mechanical strength and melting behavior, could yield valuable insights into the anomalous properties of ice. Finally, integrating these machine learning based simulation techniques with experimental studies could help validate the models further and refine our understanding of the premelting layer.

%%%%%%%%%%%%%%%%%%%%%%%%%%%%%%%%%%%%%%%%%%
\vspace{6pt} 

%%%%%%%%%%%%%%%%%%%%%%%%%%%%%%%%%%%%%%%%%%
%% optional
%\supplementary{The following supporting information can be downloaded at:  \linksupplementary{s1}, Figure S1: title; Table S1: title; Video S1: title.}


%\begin{tabularx}{\textwidth}{lll}
%\toprule
%\textbf{Name} & \textbf{Type} & \textbf{Description} \\
%\midrule
%S1 & Python script (.py) & Script of python source code used in XX \\
%S2 & Text (.txt) & Script of modelling code used to make Figure X \\
%S3 & Text (.txt) & Raw data from experiment X \\
%S4 & Video (.mp4) & Video demonstrating the hardware in use \\
%\bottomrule
%\end{tabularx}
%}

%%%%%%%%%%%%%%%%%%%%%%%%%%%%%%%%%%%%%%%%%%
\authorcontributions{For research articles with several authors, a short paragraph specifying their individual contributions must be provided. The following statements should be used ``Conceptualization, X.X. and Y.Y.; methodology, X.X.; software, X.X.; validation, X.X., Y.Y. and Z.Z.; formal analysis, X.X.; investigation, X.X.; resources, X.X.; data curation, X.X.; writing---original draft preparation, X.X.; writing---review and editing, X.X.; visualization, X.X.; supervision, X.X.; project administration, X.X.; funding acquisition, Y.Y. All authors have read and agreed to the published version of the manuscript.'', please turn to the  \href{http://img.mdpi.org/data/contributor-role-instruction.pdf}{CRediT taxonomy} for the term explanation. Authorship must be limited to those who have contributed substantially to the work~reported.}

\funding{Please add: ``This research received no external funding'' or ``This research was funded by NAME OF FUNDER grant number XXX.'' and  and ``The APC was funded by XXX''. Check carefully that the details given are accurate and use the standard spelling of funding agency names at \url{https://search.crossref.org/funding}, any errors may affect your future funding.}

\dataavailability{We encourage all authors of articles published in MDPI journals to share their research data. In this section, please provide details regarding where data supporting reported results can be found, including links to publicly archived datasets analyzed or generated during the study. Where no new data were created, or where data is unavailable due to privacy or ethical restrictions, a statement is still required. Suggested Data Availability Statements are available in section ``MDPI Research Data Policies'' at \url{https://www.mdpi.com/ethics}.} 

% Only for journal Nursing Reports
%\publicinvolvement{Please describe how the public (patients, consumers, carers) were involved in the research. Consider reporting against the GRIPP2 (Guidance for Reporting Involvement of Patients and the Public) checklist. If the public were not involved in any aspect of the research add: ``No public involvement in any aspect of this research''.}

% Only for journal Nursing Reports
%\guidelinesstandards{Please add a statement indicating which reporting guideline was used when drafting the report. For example, ``This manuscript was drafted against the XXX (the full name of reporting guidelines and citation) for XXX (type of research) research''. A complete list of reporting guidelines can be accessed via the equator network: \url{https://www.equator-network.org/}.}

\acknowledgments{In this section you can acknowledge any support given which is not covered by the author contribution or funding sections. This may include administrative and technical support, or donations in kind (e.g., materials used for experiments).}

\conflictsofinterest{The authors declare no conflicts of interest. The funders had no role in the design of the study; in the collection, analyses, or interpretation of data; in the writing of the manuscript; or in the decision to publish the results'.}

%%%%%%%%%%%%%%%%%%%%%%%%%%%%%%%%%%%%%%%%%%
\begin{adjustwidth}{-\extralength}{0cm}

\reftitle{References}


\bibliography{premelting.bib}

%=====================================
% References, variant B: internal bibliography
%=====================================

% If authors have biography, please use the format below
%\section*{Short Biography of Authors}
%\bio
%{\raisebox{-0.35cm}{\includegraphics[width=3.5cm,height=5.3cm,clip,keepaspectratio]{Definitions/author1.pdf}}}
%{\textbf{Firstname Lastname} Biography of first author}
%
%\bio
%{\raisebox{-0.35cm}{\includegraphics[width=3.5cm,height=5.3cm,clip,keepaspectratio]{Definitions/author2.jpg}}}
%{\textbf{Firstname Lastname} Biography of second author}

% For the MDPI journals use author-date citation, please follow the formatting guidelines on http://www.mdpi.com/authors/references
% To cite two works by the same author: \citeauthor{ref-journal-1a} (\citeyear{ref-journal-1a}, \citeyear{ref-journal-1b}). This produces: Whittaker (1967, 1975)
% To cite two works by the same author with specific pages: \citeauthor{ref-journal-3a} (\citeyear{ref-journal-3a}, p. 328; \citeyear{ref-journal-3b}, p.475). This produces: Wong (1999, p. 328; 2000, p. 475)

%%%%%%%%%%%%%%%%%%%%%%%%%%%%%%%%%%%%%%%%%%
%% for journal Sci
%\reviewreports{\\
%Reviewer 1 comments and authors’ response\\
%Reviewer 2 comments and authors’ response\\
%Reviewer 3 comments and authors’ response
%}
%%%%%%%%%%%%%%%%%%%%%%%%%%%%%%%%%%%%%%%%%%
\PublishersNote{}
\end{adjustwidth}
\end{document}